\documentclass[a4paper,twocolumn,11pt,accepted=2024-10-08]{quantumarticle}
\pdfoutput=1
\usepackage[utf8]{inputenc}
\usepackage[english]{babel}
\usepackage[T1]{fontenc}
\usepackage{hyperref}
\usepackage[numbers]{natbib}
\usepackage[resetlabels]{multibib}
\usepackage{listings}
\usepackage{caption}
\usepackage{bm}
\usepackage{hyperref}
\usepackage{graphicx}
\usepackage{amsmath,amsfonts,amssymb,amsthm}
\usepackage{color}
\usepackage{soul}
\usepackage{subfigure}
\usepackage{physics}
\usepackage{dsfont}
\usepackage{hyperref}
\usepackage{url}
\usepackage{leftidx}
\usepackage{textcomp}
\usepackage{gensymb}
\usepackage{threeparttable}
\usepackage{rotating}
\usepackage{esvect}
\usepackage{booktabs}
\usepackage{marvosym}
\usepackage[ruled,vlined]{algorithm2e}
\usepackage[noend]{algpseudocode}
\usepackage[dvipsnames]{xcolor}
\usepackage{mathrsfs}
\usepackage{mathtools}
\usepackage{tikz}
\usepackage{lipsum}
\newcommand{\hpppf}{\includegraphics[trim= 0 45pt 0 0, scale=0.55]{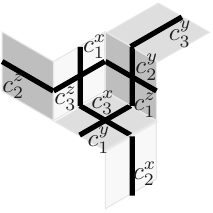}}
\newcommand{\hpppe}{\includegraphics[trim= 0 45pt 0 0, scale=0.55]{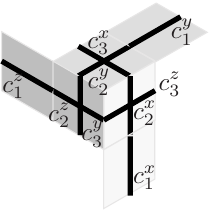}}

\newcommand{\hppmf}{\includegraphics[trim=0 45pt 0 0, scale=0.55]{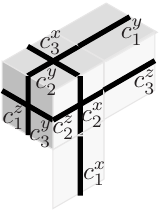}}
\newcommand{\hppme}{\includegraphics[trim=0 45pt 0 0, scale=0.55]{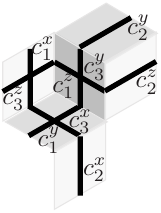}}

\newcommand{\hpmpf}{\includegraphics[trim=0 40pt 0 0, scale=0.55]{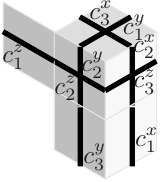}}
\newcommand{\hpmpe}{\includegraphics[trim=0 40pt 0 0, scale=0.55]{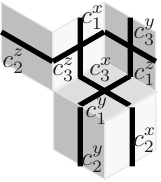}}

\newcommand{\hpmmf}{\includegraphics[trim= 0 40pt 0 0, scale=0.55]{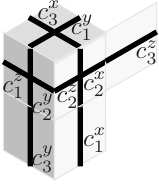}}
\newcommand{\hpmme}{\includegraphics[trim= 0 40pt 0 0, scale=0.55]{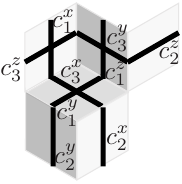}}

\newcommand{\hmppf}{\includegraphics[trim=0 32pt 0 0, scale=0.55]{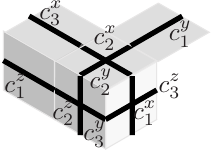}}
\newcommand{\hmppe}{\includegraphics[trim=0 32pt 0 0, scale=0.55]{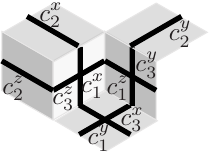}}

\newcommand{\hmpmf}{\includegraphics[trim=0 32pt 0 0, scale=0.55]{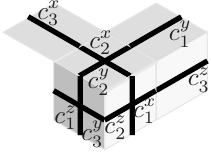}}
\newcommand{\hmpme}{\includegraphics[trim=0 32pt 0 0, scale=0.55]{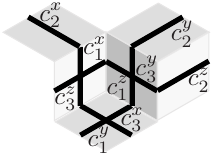}}

\newcommand{\hmmpf}{\includegraphics[trim=0 45pt 0 0, scale=0.55]{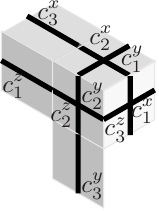}}
\newcommand{\hmmpe}{\includegraphics[trim=0 45pt 0 0, scale=0.55]{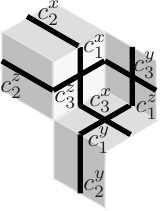}}

\newcommand{\hmmmf}{\includegraphics[trim= 0 45pt 0 0, scale=0.55]{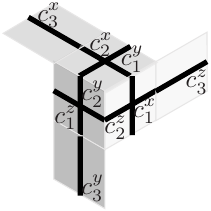}}
\newcommand{\hmmme}{\includegraphics[trim= 0 45pt 0 0, scale=0.55]{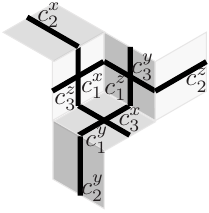}}

\newcommand{\humxbr}{\includegraphics[trim= 0 25pt 0 0, scale=0.4]{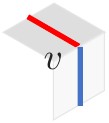}}
\newcommand{\humybr}{\includegraphics[trim= 0 25pt 0 0, scale=0.4]{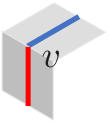}}
\newcommand{\humzbr}{\includegraphics[trim= 0 20pt 0 0, scale=0.4]{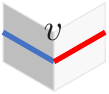}}

\newcommand{\hbdrxy}{\includegraphics[trim= 0 10pt 0 0, scale=0.5]{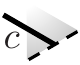}}
\newcommand{\hbdryz}{\includegraphics[trim= 0 20pt 0 0, scale=0.5]{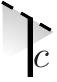}}
\newcommand{\hbdrzx}{\includegraphics[trim= 0 10pt 0 0, scale=0.5]{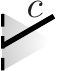}}
\newcommand{\hbdrxz}{\includegraphics[trim= 0 20pt 0 0, scale=0.5]{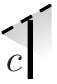}}
\newcommand{\hbdrzy}{\includegraphics[trim= 0 10pt 0 0, scale=0.5]{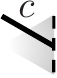}}
\newcommand{\hbdryx}{\includegraphics[trim= 0 10pt 0 0, scale=0.5]{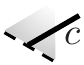}}

\begin{document}

\title{Quantum lozenge tiling and entanglement phase transition}

\author{Zhao Zhang}
\affiliation{Department of Physics, University of Oslo, P.O. Box 1048 Blindern, N-0316 Oslo, Norway}
\affiliation{SISSA and INFN, Sezione di Trieste, via Bonomea 265, I-34136, Trieste, Italy}
\orcid{0000-0002-9425-732X}
\email{zhaoz@uio.no}
\author{Israel Klich}
\affiliation{Department of Physics, University of Virginia, Charlottesville, VA, USA}
\orcid{0000-0002-8979-0170}

\maketitle

\begin{abstract}
While volume violation of area law has been exhibited in several quantum spin chains, the construction of a corresponding ground state in higher dimensions, entangled in more than one direction, has been an open problem. Here we construct a 2D frustration-free Hamiltonian with maximal violation of the area law. We do so by building a quantum model of random surfaces with color degree of freedom that can be viewed as a collection of colored Dyck paths. The Hamiltonian may be viewed as a 2D generalization of the Fredkin spin chain. It relates all the colored random surface configurations subject to a Dirichlet boundary condition and hard wall constraint from below to one another, and the ground state is therefore a superposition of all such classical states and non-degenerate. Its entanglement entropy between subsystems undergoes a quantum phase transition as the deformation parameter is tuned. The area- and volume-law phases are similar to the one-dimensional model, while the critical point scales with the linear size of the system $L$ as $L\log L$. Further it is conjectured that similar models with entanglement phase transitions can be built in higher dimensions with even softer area law violations at the critical point.

\end{abstract}

%
\section{Introduction}
\label{sec:Intro}

Entanglement entropy (EE) and its scaling has been a central theme of quantum many-body physics, not only because entanglement is a unique feature in the quantum world by itself, but also for their crucial role in determining the computational complexity of the numerical simulations of quantum many-body systems, indication of topological order and understanding of the holographic principle and black hole entropy. While EE of a generic eigenstate in the Hilbert space is shown to scale with the systems size~\cite{Page93}, EE of the ground states of gapped local Hamiltonians are generally observed to obey the so-called area law, scaling with the size of the boundary. A milestone in the study of area-law has been Hastings' rigorous proof of the result in one-dimensional systems~\cite{Hastings_2007}. Recently, a similar result in two-dimension has been proven for frustration-free models~\cite{AnshuEA21}. While area-law has been ubiquitous in gapped systems, plenty of examples of area-law violation has also been found in various gapless systems. (1+1)-dimensional critical system described by a conformal field theory has EE of  logarithmic scaling~\cite{Calabrese_2009}. Entanglement entropy of free fermions with a Fermi sea in dimension $d$ scales as $L^{d-1}\log{L}$~\cite{GioevEA06}. In one dimension, lattice models with even more severe violation has also been found in a class of frustration-free (FF) models dubbed Motzkin and Fredkin spin chains with up to volume-law scaling of EE~\cite{BravyiEA12, Movassagh13278, Zhang5142, PhysRevB.94.155140, Salberger:2017aa, Salberger_2017, Zhang_2017}. Despite the plethora of 1D lattice models with such violations, their 2D counterparts are yet to be discovered. In this manuscript, we give the first successful example of such a model with exotic entanglement scaling on a two-dimensional lattice.

A crucial ingredient of the Motzkin and Fredkin models is that, by employing next-nearest-neighbor interactions and boundary conditions, they allow a well-defined height representation that carries long-range entanglement which local spins could not. The first difficulty to generalize this to two dimension is to make sure the height function does not give rise to any ambiguity when counting around a closed loop in the lattice. This problem is intrinsically avoided in the U(1) Coulomb gas phase, naturally emerged in constrained Hilbert space of fully packed dimer or loop covering models~\cite{ArdonneEA04, Kenyon:2008wf, Roising22}. In Ref.~\cite{Roising22}, a first attempt was made to build two models using fully packed loops on a square lattice, but the entanglement entropy computed there obeys area law, as the ground state is a projected entangled pair state contracted by the local constraints in Hilbert space. In the combinatorics and classical statistical mechanics literature, the fully packed dimer configurations of a honeycomb lattice are mapped to random tilings of lozenges, and are extensively studied in the context of limit shape behavior and arctic curves~\cite{Gorin:2021vd}, and are recently used to study two dimensional Kardar–Parisi–Zhang (KPZ) growth~\cite{Borodin:2018aa}. Motivated by the resemblance between the Fredkin spin chains and the models of asymmetric simple exclusion process (ASEP), it is promising that a quantum version of lozenge tiling could lead to exotic scalings of the height function that facilitate area violation of EE. A 2D generalization to the Fredkin moves between lozenge tilings was first proposed in Ref.~\cite{Lamacraft22} as a classical stochastic model. In this manuscript, we incorporate those dynamics into a quantum Hamiltonian with projection operators that annihilate a ground state of uniform superposition of lozenge tilings. In principle, one could compute the bipartite entanglement entropy of such a quantum model using the information of the limit shape across the interface between the two subsystems and the fluctuation around it to compute the entanglement entropy scaling in the thermodynamic limit. Moreover, one can also perform q-deformations to the Hamiltonian and obtain entanglement entropy of q-weighted superposition from results in q-enumerations of lozenge tilings~\cite{qLozenge}.

A yet richer model can be defined when the local Hilbert space is enlarged by introducing a color degree of freedom to the dimers. Two- and multi-color dimer models has been studied both on the square lattice~\cite{Raghavan:1997vi} and honeycomb lattice~\cite{MulticolorDimer}. Color-code models have also been heavily studied in the context of exactly solvable model of topological order~\cite{ColorCode}, as generalizations of Kitaev's toric code \cite{Zhang2024bicolorloopmodels}, topological defect~\cite{ColorcodeTeo}, and as a candidate for fault-tolerant quantum computation~\cite{Kesselring2018boundariestwist}. Recently, color degrees of freedom have been used to generalize the extensively entangled rainbow chain \cite{Vitagliano_2010,Ramirez_2014} to a quasi-1D lattice embedded in 2D space with inhomogeneous coupling strength decaying from the center \cite{ZHANG2023169395}.
For 1D spin chains, the dependency of entanglement entropy scaling on the number of local degrees of freedom has been studied with an N-component partially integrable chain~\cite{PhysRevB.106.134420}, which can be viewed as a low-energy effective Hamiltonian of a related quasi-2D spin ladder model~\cite{PhysRevLett.125.240603}. In particular, the entanglement entropy of integrable excited states was shown to decompose into a part that scales only with the size of the system and one that scales also with the number of the dimensionality of local Hilbert space. Perhaps the power of color degrees of freedom in drastically changing the scaling behavior of entanglement entropy is only fully unleashed when they are facilitated by a notion of the height variable, as has been manifested in the Motzkin and Fredkin spin chains. In both those cases and the present model, EE also decomposes into a contribution from color degrees of freedom and the fluctuating height degree of freedom. The leading contribution comes from the entanglement among color degrees of freedom, but its scaling behavior is determined by the average shape of the random surface described by the height function, which can be deformed continuously and undergoes a phase transition between different scaling behaviors across a critical point.

The rest of the paper is organized as follows. In Sec.~\ref{sec:Model}, we define the lattice, Hilbert space and Hamiltonian of the model of quantum lozenge tiling, and show that its unique ground state is a superposition of random surfaces subject to a hard wall constraint from below. In Sec.~\ref{sec:Entropy}, the ground state EE is shown to decompose into contributions from the color and height degrees of freedom separately, and the former is proportional to the average cross-sectional area underneath the random surface. Sec.~\ref{sec:MFT} takes the scaling limit of the height function and provides intuitive arguments for various scaling behaviours of EE for different deformation parameter. Rigorous mathematical theorems on the critical scaling behavior was referenced to establish the existence of an entanglement phase transition. Sec.~\ref{sec:higherd} briefly discusses the analogous but quantitatively different EE scaling of ground states at critical point of quantum tiling models in higher dimensions. Sec.~\ref{sec:Concl} gives a conclusion as well as several possible future directions to pursue.

%
\section{Quantum lozenge tiling}
\label{sec:Model}
%
\begin{figure}[t!bh]
	\centering
	\includegraphics[width=0.8\linewidth]{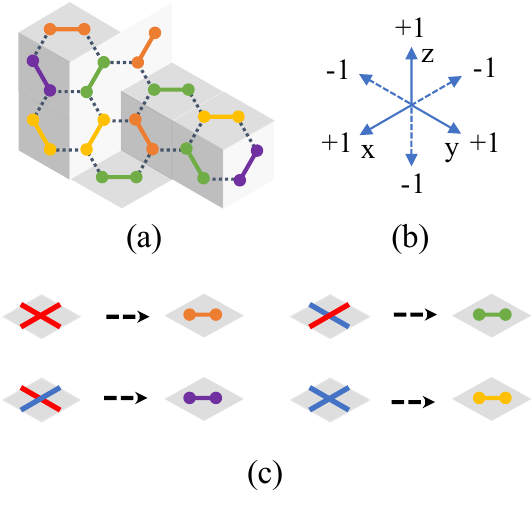} 
	\caption{(a) Mapping from $s^2$-colored dimer coverings on honeycomb lattice to lozenge tilings with an $s$-colored line along each pair of parallel sides. (b) Convention of height change between plaquette centers and vertices of lozenges: along the positive direction of x,y,z axes height increases by 1. (c) The $s^2$-coloring of dimers can be mapped from a two-component $s$-coloring of lines, one directed along each edge of the lozenge. The mapping for dimers oriented along the other two direction can be obtained by rotations.}
	\label{fig:conven}
\end{figure}

\begin{figure}[t!bh]
	\centering
	\includegraphics[width=0.8\linewidth]{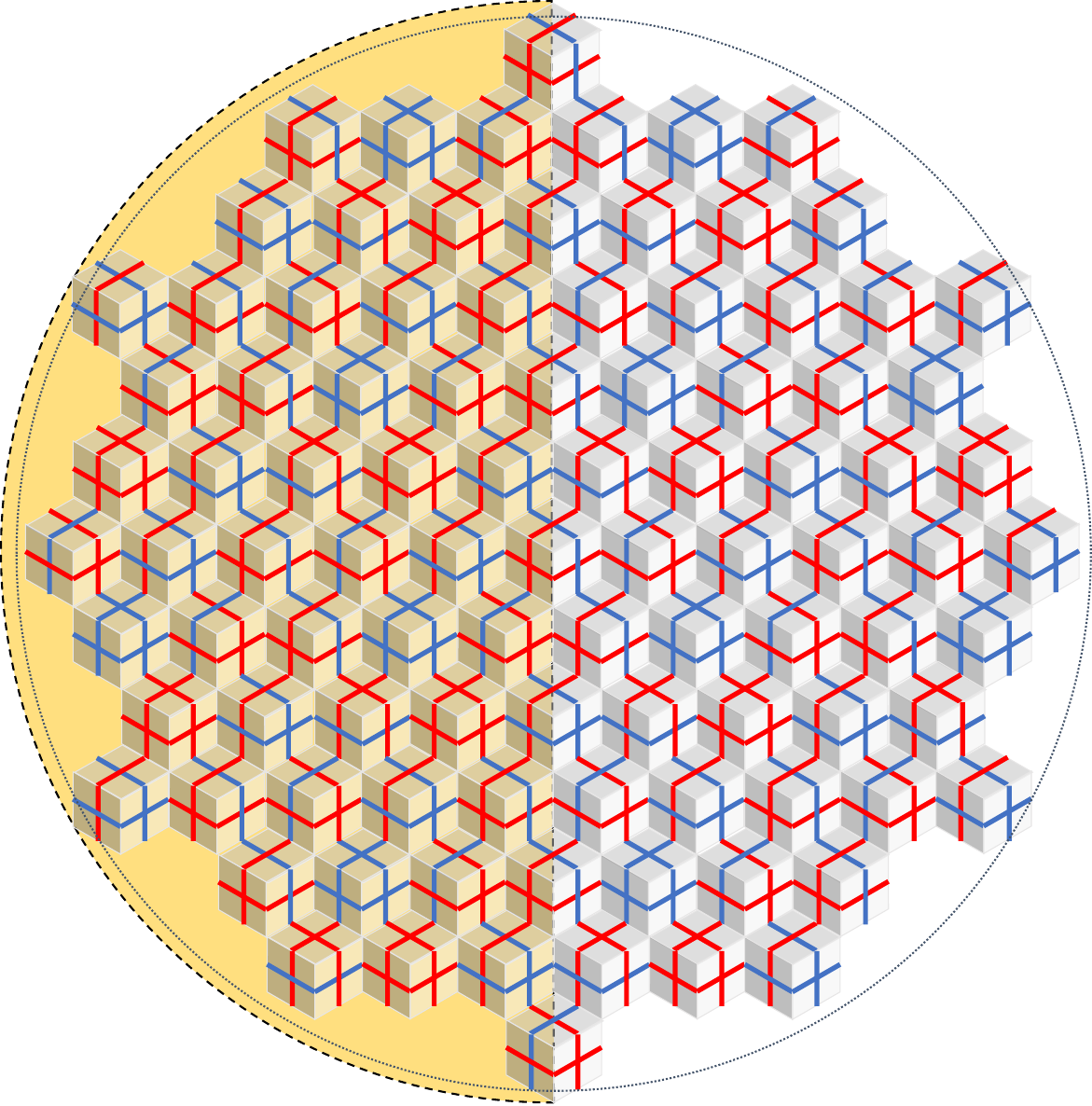} 
	\caption{A coloring of the minimal volume tiling configuration of a lattice of linear size $L=20$. The lattice resides within the arctic circle of the usual lozenge tiling of a hexagonal boundary, where the height can fluctuate. In the scaling limit, the boundary of the lattice approaches the arctic circle, and has constant height $0$ along the boundary, where as in the discrete case, the height fluctuates between $\pm \frac{1}{2}$. The dashed line marks the cut into the shaded and unshaded subsystems. The three-dimensional effect is understood with light shining against the y-axis slightly tilted downward. This particular tiling is also known as the rhombille tiling in the mathematical literature. }
	\label{fig:Mintiling}
\end{figure}

\begin{figure}[t!bh]
	\centering
	\includegraphics[width=\linewidth]{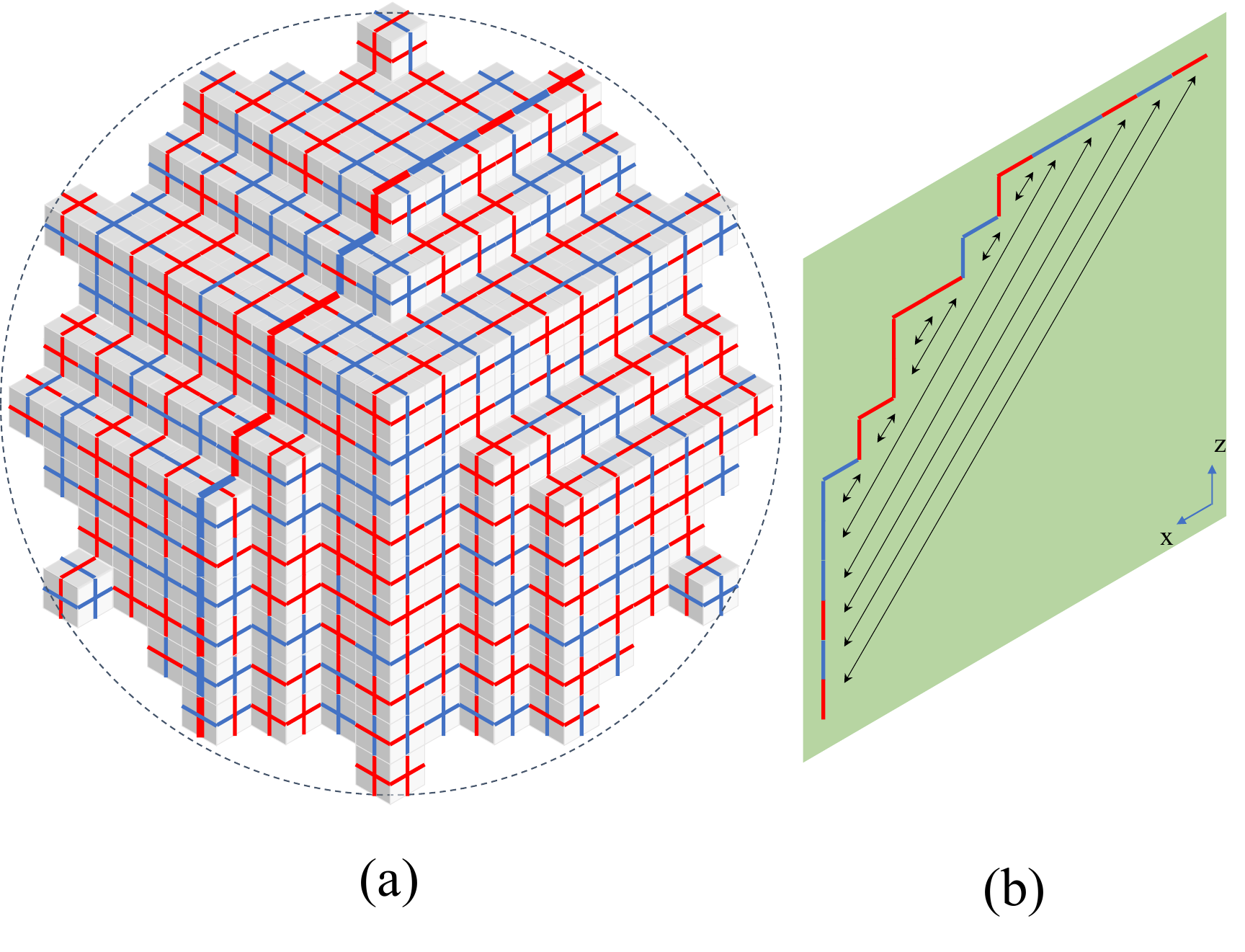} 
	\caption{(a) A coloring of the maximal volume tiling configuration of a lattice of linear size $L=20$. In the thermodynamic limit, the surface approaches a dome shape. (b) One slice of the x-z plane which intersects with the maximal surface in (a) giving a colored Dyck path. The corresponding path in (a) is thickened.}
	\label{fig:Maxtiling}
\end{figure}

Lozenge tiling can be viewed as a covering of the faces of triangular lattice, where each tile occupies two adjacent triangular faces and each triangular face of the lattice is covered by one lozenge and one lozenge only. We call this triangular lattice $\Lambda$, and its dual, hexagonal lattice $\Lambda^*$. Each face of the triangular lattice maps to a vertex in the dual honeycomb lattice, and a lozenge tiling therefore maps to a dimer covering of the hexagonal lattice. Our model further allows the lozenge tiles or equivalently dimers on the hexagonal lattice to come in $s^2$ different possible colors, as is shown in Fig.~\ref{fig:conven} (a). 
On each edge of the honeycomb lattice, the local degrees of freedom are either uncovered, or covered by a dimer in one of the $s^2$ internal states. Furthermore, the global Hilbert space is constrained by the fully packed dimer covering condition, which means that each vertex is covered by one dimer and one dimer only. The constraint can be realized by the Hamiltonian 
\begin{equation}
    H_0=\sum_{\boldsymbol{r}\in \Lambda^*}(\sum_{i=1}^3 n_{\boldsymbol{r},\boldsymbol{r}+\boldsymbol{e}_i}-1)^2,
\end{equation}
where $n_{\boldsymbol{r},\boldsymbol{r}+\boldsymbol{e}_i}$ is the number operator of the dimer living on the edge along the $\boldsymbol{e}_i$ direction at site $\boldsymbol{r}$. 
The constraint of fully packed dimers allows a well-defined height function on the dual triangular lattice. We adopt the height change convention as depicted in Fig.~\ref{fig:conven} (b), namely the height increases by 1 along the positive directions of the three axis and decreases by 1 in the opposite directions, when counting along the edge of a lozenge. Accordingly the height changes by $\pm 2$ across the shorter diagonal.

A theorem by Thurston~\cite{Thurstontileability} says that for a triangular lattice to be tileable by lozenges, the difference between the height of two vertices on the boundary defined by the convention in Fig.~\ref{fig:conven} (b) is bounded by the minimal number of edges in a positively oriented path connecting the two vertices. The lattice of our model satisfies the stronger constraint that the difference in height along the boundary is either $0$ or $1$, by requiring consecutive edges along the boundary to rotate by an angle of $\pm \frac{\pi}{3}$. As a result, the number of three types of lozenges in the tiling, or the number of dimers with three different orientation are equal in our lattice. In Fig.~\ref{fig:Mintiling} and \ref{fig:Maxtiling} (a), we show two tiling configurations with a boundary inside a circle of diameter $L=20$. The different shading of lozenges in three different orientations reveals an intuitive 3D bird's-eye view of a pile of unit cubes, with the former configuration referred to as the ``rhombille tiling'', while the latter resembles a tetrahedron. As the lattice spacing decreases, the boundary of our lattice approaches the circle by filling as many hexagons in the empty space inside the circle as possible, so that in the thermodynamic or scaling limit, the lattice has a higher isotropy than the three-fold rotation symmetry. The particular choice of shape for the lattice boundary is motivated by the uniqueness of the ground state, which will become clear shortly. 

 The dimers in our Hilbert space have an internal degree of freedom of $s^2$ states mapping to a two-component coloring of the lozenges. We denote the two components with colored lines along the two edges in the middle, with each component having choices of $s$ colors. (See Fig.~\ref{fig:conven} (c).) Since the height variable is defined up to a global constant shift, we further specify from now on that the height at the midpoint of the boundary edges of the lattice, i.e.~the endpoints of the colored lines, to be $0$, so that the height at any other vertex of the triangular lattice can be uniquely defined according to the convention in Fig.~\ref{fig:conven} (b).

 A coloring of the minimal height configuration is shown in Fig.~\ref{fig:Mintiling} (here the height in the bulk alternate between 0 and 1). A coloring of the maximal height configuration is depicted in Fig.~\ref{fig:Maxtiling}. The notion of height can be interpreted as the distance to the (1,1,1)-plane amplified by a factor of $\sqrt{3}$, if we view the tiling as a 3D pile of unit cubes, taking lozenges in three different orientation as facets projected to the x-y, y-z, and z-x plane respectively. 

 Tracing each of the lines in Figs.~\ref{fig:Mintiling} and \ref{fig:Maxtiling} within a plane perpendicular to the $x-, y-$ or $z-$axes, we see a Dyck path (Fig.~\ref{fig:Maxtiling}(b)), a path of up/down moves that starts and ends at the same height and never goes below the initial height. While each dimer covering can have arbitrary coloring along such a path, we introduce a Hamiltonian that enforces the ground state to have only $s-$coloring that correspond to a proper coloring as in the Motzkin and Fredkin spin chains. These correlated color degrees of freedom will be the main source of entanglement in the system.

We now construct a Hamiltonian the ground state of which is a weighted superposition of such tessellations, with proper coloring. The dynamics that takes one tiling configuration to another locally different one is introduced with local interactions relating the eight pairs of configurations in \eqref{eq:FredkinMs}, reminiscent of the Fredkin chain, and diagonal boundary terms that penalize configurations with height below 0 right inside the boundary. The latter takes the form
\begin{equation}
\begin{split}
    H_\partial = \sum_{\mathrm{edges}\in \partial \Lambda}&\sum_{c}\Big(\ket{\hbdrzy}\bra{\hbdrzy}+\ket{\hbdrzx}\bra{\hbdrzx}\\
    +&\ket{\hbdrxy}\bra{\hbdrxy}+\ket{\hbdrxz}\bra{\hbdrxz}\\
    +&\ket{\hbdryz}\bra{\hbdryz}+\ket{\hbdryx}\bra{\hbdryx}\Big),
\end{split}
\label{eq:Hbdr}
\end{equation}
where $c$ denotes the color of the line attached to the boundary, and we used dashed line to denote the boundary of the lattice and the shaded triangle to denote the interior side of the boundary. The interaction appears in the Hamiltonian as projection operators requiring the ground state to have weighted superposition of these pairs, given by
\begin{equation}
\begin{split}
	H_T=\sum_{p\in \Lambda^\circ}&\sum_{\{\bm{c}^x,\bm{c}^y,\bm{c}^z\}} \sum_{f_{x,y,z}=\pm}\\ &\frac{1}{[2]_{q^3}}\ket{T_{f_x,f_y,f_z}^{\bm{c}^x,\bm{c}^y,\bm{c}^z}}_p\bra{T_{f_x,f_y,f_z}^{\bm{c}^x,\bm{c}^y,\bm{c}^z}}_p ,
\end{split}	
\label{eq:HT}
\end{equation}
where $[2]_q\vcentcolon=q+q^{-1}$ normalizes the projector in the presence of the deformation parameter $q$, and $\ket{T_{f_x,f_y,f_z}^{\bm{c}^x,\bm{c}^y,\bm{c}^z}}$ as detailed in Eq.~\eqref{eq:FredkinMs}. The first summation in 
\eqref{eq:HT} is over all 6-lozenge neighborhood $p$'s in the bulk or interior of lattice, denoted by $\Lambda^\circ$, and each of the three components of the color vectors $\bm{c}^x,\bm{c}^y,\bm{c}^z$, which always refer to the edge above (resp. on the left) for the $y,z$ (resp. $x,y$) components, can take $s$ different values. 
\begin{equation}
\begin{aligned}
	\ket{T_{+,+,+}^{\bm{c}^x,\bm{c}^y,\bm{c}^z}\!}&\!=\! q^{\!-\!\frac{3}{2}}\!\ket{\hpppf\!}\! -\! q^{\frac{3}{2}}\!\ket{\hpppe\!}\!,\\
	\ket{T_{+,+,-}^{\bm{c}^x,\bm{c}^y,\bm{c}^z}\!}&\!=\! q^{\!-\!\frac{3}{2}}\!\ket{\hppmf}\! -\!q^{\frac{3}{2}}\!\ket{\hppme}\!,\\
    \ket{T_{+,-,+}^{\bm{c}^x,\bm{c}^y,\bm{c}^z}\!}&\!=\!q^{\!-\!\frac{3}{2}}\!\ket{\hpmpf} \! -\! q^{\frac{3}{2}}\!\ket{\hpmpe}\!,\\
	\ket{T_{+,-,-}^{\bm{c}^x,\bm{c}^y,\bm{c}^z}\!}&\!=\!q^{\!-\!\frac{3}{2}}\!\ket{\hpmmf} \! -\!q^{\frac{3}{2}}\!\ket{\hpmme}\!,\\
	\ket{T_{-,+,+}^{\bm{c}^x,\bm{c}^y,\bm{c}^z}\!}&\!=\!q^{\!-\!\frac{3}{2}}\!\ket{\hmppf} \! -\! q^{\frac{3}{2}}\!\ket{\hmppe}\!,\\
	\ket{T_{-,+,-}^{\bm{c}^x,\bm{c}^y,\bm{c}^z}\!}&\!=\!q^{\!-\!\frac{3}{2}}\!\ket{\hmpmf} \! -\! q^{\frac{3}{2}}\!\ket{\hmpme}\!,\\
	\ket{T_{-,-,+}^{\bm{c}^x,\bm{c}^y,\bm{c}^z}\!}&\!=\!q^{\!-\!\frac{3}{2}}\!\ket{\hmmpf} \! -\! q^{\frac{3}{2}}\!\ket{\hmmpe}\!,\\
	\ket{T_{-,-,-}^{\bm{c}^x,\bm{c}^y,\bm{c}^z}\!}&\!=\!q^{\!-\!\frac{3}{2}}\!\ket{\hmmmf} \! -\! q^{\frac{3}{2}}\!\ket{\hmmme}\!.
\end{aligned}
\label{eq:FredkinMs}
\end{equation}
Note that the projectors in Eq. \eqref{eq:HT} relate between tilings where the minimal height in the neighbors remain at the boundary of the neighborhood. Therefore applying the projectors cannot make the height negative in the bulk of the lattice as long as the boundary is kept at zero height.

Note that none of the vertices in Fig.~\ref{fig:Mintiling} overlaps with the local configurations in the first column of \eqref{eq:FredkinMs}. So as long as the height of the lozenges along the boundary is set to be 0 and increases going towards inside the bulk, which is done by the boundary Hamiltonian $H_\partial$, the height configuration of the entire lattice are lowered bounded by 0. That boundary Hamiltonian can be written as a sum of local projection operators that penalizes local boundary configurations different from those shown in Fig.~\ref{fig:Mintiling} and \ref{fig:Maxtiling}. 

The kinetically constrained bulk Hamiltonian could fragment the space of non-negative height configurations satisfying the Fredkin condition into Krylov subspaces, but an ergodicity argument given in the  Appendix~\ref{sec:ergo} shows that all such configurations participate in the ground state superposition. So far, all the configurations between the minimal height configuration of rhombille tiling, and the maximal height configuration of a tetrahedron are superposed to form a ground state superposition weighted by the volume underneath. Yet different colorings of lozenges in the same height configuration still live in distinct Krylov sectors. To obtain a unique ground state as a superposition of different colorings, we add color mixing terms in the Hamiltonian that require the components of color perpendicular to the convex corners of edges where two differently oriented lozenges meet to be matched and either color to come in equal probability:
\begin{equation}
\begin{split}
	H_\Gamma=&\sum_{v\in \Lambda^\circ}\sum_{n=x,y,z}\sum_{c_1\ne c_2}\ket{P_{v,n}^{c_1,c_2}}\bra{P_{v,n}^{c_1,c_2}}+\\
	&\frac{1}{2}\left(\!\ket{P_{v,n}^{c_1,c_1}}\! -\!\ket{P_{v,n}^{c_2,c_2}}\! \right)\!\left(\!\bra{P_{v,n}^{c_1,c_1}} \!-\!\bra{P_{v,n}^{c_2,c_2}}\right)\!	,			
\end{split}
\end{equation}
where, e.g.
\begin{equation*}
	\ket{P_{v,x}^{\mathrm{b,r}}}\!=\!\ket{\humxbr}\!,
    \ket{P_{v,y}^{\mathrm{b,r}}}\!=\!\ket{\humybr}\!,
	\ket{P_{v,z}^{\mathrm{b,r}}}\!=\!\ket{\humzbr}\!.\!
\end{equation*}
The full Hamiltonian is the sum of all the terms introduced above
\begin{equation}
	H=H_0+H_\partial+H_T+H_\Gamma.
\end{equation}
It is frustration free as all the terms share the common ground state 
\begin{equation}
	\ket{\mathrm{GS}}=\frac{1}{\sqrt{\mathcal{N}}}\sum_{h(\partial T)=0}\prod_{v\in \Lambda^\circ}\theta(h_v)\sum_{C}{}^{'}q^{\mathcal{V}(T)}\ket{T^C},
\label{eq:GS}
\end{equation}
where $\mathcal{N}$ is the normalization constant that depends only on lattice size $L$ and deformation parameter $q$, the first sum denotes summation over tiling configurations with a Dirichlet boundary condition on the height function, subject to the product of Heaviside step function $\theta$ of the height of vertices in the bulk. The primed sum denotes summation over coloring configurations that along any cross-sectional paths in the x-y, y-z and z-x plane, the lines of lozenges match in color with their partner of the same height, and zero otherwise. The volume of a surface  $\mathcal{V}(T)=\sum_{v\in \Lambda^\circ}h_v(T)$ is defined by the sum of the height of all the vertices in the bulk of the lattice. 
\section{Entanglement of colored surfaces}
\label{sec:Entropy}
\begin{figure}[t!bh]
	\centering
	\includegraphics[width=\linewidth]{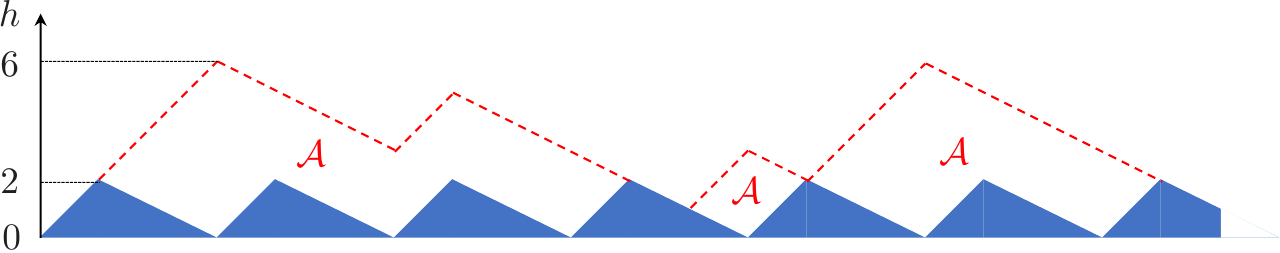} 
	\caption{Cross-sectional view of the random faceted dome surface (red dashed line). The cross-sectional area $\mathcal{A}$ is defined as the sum of the height difference between the red dashed line and the blue zigzag denoting the lowest height surface in Fig.~\ref{fig:Mintiling}.}
	\label{fig:crosssec}
\end{figure}

Choosing a vertical cut across the center, the Schmidt decomposition of the ground state \eqref{eq:GS} can be written as
\begin{equation}
    \ket{\mathrm{GS}}=\sum_{\vec{h};\vec{c}}\sqrt{\frac{M_{\vec{h}}^2}{\mathcal{N}}}\ket{T^{\vec{c}}_{\vec{h}}}_{\mathrm{L}}\otimes\ket{T^{\vec{c}}_{\vec{h}}}_{\mathrm{R}},
\end{equation}
where $\vec{c}$ denotes all of the colors of the edges to be matched with another edge on the other side of the bipartition, and 
\begin{equation}
	\ket{T^{\vec{c}}_{\vec{h}}}_{\mathrm{L(R)}}=\frac{1}{\sqrt{M_{\vec{h}}}}\sideset{}{'}\sum q^{\mathcal{V}_{\mathrm{L(R)}}(T)}\ket{T^C}_{\mathrm{L(R)}}
\end{equation}
are normalized wave functions of the left (resp.~right) subsystems, including half horizontal lozenges wherever the cut intersects them, and the primed sum is a shorthand notation for summing over tiling configurations in the left (resp. right) subsystem with height profile $\vec{h}$ on the middle boundary and coloring $\vec{c}$ of the dangling lozenges with one or two of their color components entangled in the other subsystem.\footnote{Since the entanglement cut goes through the horizontal lozenges in the middle, when defining the ``half system'' states $\ket{T^C}_{\mathrm{L(R)}}$, we include half of the horizontally oriented lozenge in the left subsystem, and half in the right subsystem. So the fact that they are from the same lozenge sharing the same color implies that it also contributes to the entanglement. The weight of the colored tiling configurations $\ket{T^C}_{\mathrm{L(R)}}$ in the superposition depends only on the volume underneath the left (resp. right) part of the random surface $\mathcal{V}_{\mathrm{L(R)}}(T)$, which is equal to the sum of the height variable over all the lattice sites of left (resp. right) sublattice.} The total number of entangled pairs of color components $N_C$ depends only on the size of the lattice, not on the particular tiling configuration, whereas the unmatched color pairs is fixed by the height profile along the cut. The latter can be determined by noticing that raising a local height by any of the eight moves in \eqref{eq:FredkinMs}, the number of entangled pairs across the cut increases by 3, one from the pair of color matching lines in the x-y plane of vertical lozenges, and two from the two lines in x and y direction of the horizontal lozenge that left and right subsystem share. This amounts to $\mathcal{A}(\vec{h})+L+1$~\footnote{This is the value specific to the boundary in Fig.~\ref{fig:Mintiling}. For a generic boundary, it should be $\mathcal{A}(\vec{h})+3\lceil \frac{L}{3}\rceil$}, with $\mathcal{A}(\vec{h})$ defined as the cross-sectional area under the random surface in the cross-section of the cut, as shown in Fig.~\ref{fig:crosssec}, i.e. the sum over heights along the cut minus that of the minimal height surface. So we have
\begin{equation}
	M^2_{\vec{h}}=s^{N_C-\mathcal{A}(\vec{h})-L-1}\sideset{}{'}\sum q^{2\mathcal{V}(T)},
\end{equation}
where as before the prime indicates that the sum is over colorless tiling $T$ with height $\vec{h}$ along the cut, and
\begin{equation}
	\mathcal{N}=\sum_{\vec{h}}s^{\mathcal{A}(\vec{h})+L+1}M^2_{\vec{h}}\equiv s^{N_C}\sum_{T}q^{2\mathcal{V}(T)}.
\end{equation}
The Schmidt coefficients
\begin{equation}
	p_{\vec{h},\vec{c}}=\frac{M^2_{\vec{h}}}{\mathcal{N}}=s^{-\mathcal{A}(\vec{h})-L-1}p_{\vec{h}},
\end{equation}
can be factorized into the product of a constant conditional probability of having a particular coloring of the dangling pairs matched in color across the cut, and the marginal probability $p_{\vec{h}}\equiv\sum_{\vec{c}}p_{\vec{h},\vec{c}}$ for an uncolored random surface to have height profile $\vec{h}$ along the cut.
This results in a decomposition of the entanglement entropy into contributions from color and height degrees of freedom:
\begin{equation}
\begin{aligned}
	S_L(q,s)=&-\sum_{\vec{h};\vec{c}} p_{\vec{h},\vec{c}} \log p_{\vec{h},\vec{c}}\\ =&-\sum_{\vec{h}}  p_{\vec{h}}\log p_{\vec{h},\vec{c}}\\
    =&-\sum_{\vec{h}}p_{\vec{h}}\log s^{-\mathcal{A}(\vec{h})-L-1}-\sum_{\vec{h}}  p_{\vec{h}}\log p_{\vec{h}}\\
	=&\log s\left(\langle\mathcal{A}\rangle_{\vec{h}}+L+1\right)+S_L(q,s=1).
\label{eq:Entropy}
\end{aligned}
\end{equation} 
The former is given in terms of the average cross-sectional area under the intersection of an uncolored random surface and the bipartition plane $\langle\mathcal{A}\rangle_{\vec{h}}$ (averaged over different cross-sectional height configurations $\vec{h}$). The latter is a contribution $S_L(q,s=1)$ from the fluctuation of the shape of uncolored random domes. We notice in passing that the bipartite entanglement for cuts at $\frac{2\pi}{3}$ angle is exactly the same as that of the vertical cut. For more generic bipartition passing through the center of the system, the entanglement entropy will differ quantitatively, but share the scaling property.

\section{Scaling limit and phase transition}
\label{sec:MFT}

To evaluate the scaling property of the leading contribution $\langle\mathcal{A}\rangle_{\vec{h}}$ in \eqref{eq:Entropy}, we can use the normalization $\frac{\mathcal{N}}{s^{N_C}}$ as a partition function for the height function, as was previously applied in the study of the dynamics of the one-dimensional Motzkin and Fredkin chains~\cite{Chen:2017aa,PhysRevB.96.180402}. A continuous field of the height configuration can be defined as a piece-wise linear function $h(\bm{x})$, which takes the value of $h_v$ at the coordinates of the vertices. 
The scaling limit of lozenge tiling, where the size of each lozenge becomes infinitesimal, is described by a smooth random surface subject to a Gibbs measure that penalizes spatial variation of height with a surface tension depending on the height gradient~\cite{Gorin:2021vd,cohn2001variational,KenyonOkounkov, Destainville}. With a generic boundary shape, the height gradient along the boundary typically would freeze the height configuration near some corners of the tiling lattice, leaving only the region inside what is called an ``arctic curve'' fluctuating in height. Our lattice has been carefully defined with a constant height $0$ along the boundary, which would otherwise be, for instance, the arctic circle of a hexagonal boundary, so that the entire bulk of system fluctuates in height, with a Dirichlet boundary condition of $h|_{\partial\Lambda} =0$. The boundary Hamiltonian \eqref{eq:Hbdr} further constrains that the height in the immediate interior of the lattice boundary can only go in one direction, together with the bulk Hamiltonian \eqref{eq:HT}, they ensures that the height in the bulk of the lattice can only be non-negative. In the $q$-deformed ground state, the random surfaces are further weighted by the volume enclosed between the fluctuating surface and a flat hard wall at $h=0$, which can be accounted for by a linear potential energy in addition to the surface tension that depends on the derivative of the height function. It is the competition between the ``entropy'' of random surface captured by the surface tension $\sigma (\bm{\nabla} h(\bm{x}))$, and its ``energy'' arising from $q$-deformation, that drives the phase transition between phases with different scaling properties of the height function. Taking into account all of the above discussion,  the partition function for the shape of the random surface is given by
\begin{equation}
	Z=\int\mathcal{D}h(\bm{x})e^{\int d^2\bm{x}\big(-\sigma(\bm{\nabla} h(\bm{x}))+2(\log q)  h(\bm{x})\big)},
\label{eq:partf}
\end{equation}
where $\mathcal{D}h(\bm{x})$ is a continuous version of $\prod_v \int_0^{+\infty} dh_v\equiv \prod_v \int_{-\infty}^{+\infty}dh_v\theta(h_v)$. Notice that the discreteness of the original problem should be reflected in the requirement that the height variable $h(\bm{x})$ obeys the Lipschitz condition $|\bm{\nabla}h(\bm{x})|\le 1$, which implies the upper bound of the integration should be finite and grow linearly with the distance from the boundary of the lattice. However, as we will soon see, without the hard wall constraint, the random surfaces behaves as Gaussian free field, the large deviation of which will be exponentially improbable anyway. So if would not make a difference to relax the upper bound to $+\infty$.
The surface tension is bounded (by the entropy density of dimer coverings of the Honeycomb lattice) and can be Taylor expanded as
\begin{equation}
	\sigma(\bm{\nabla} h)=\sigma_{11}(\partial_1 h)^2+\sigma_{12}\partial_1 h \partial_2 h+ \sigma_{22}(\partial_2 h)^2 +\cdots. 
\end{equation}
The absence of odd order terms is clear from symmetry considerations, while higher order terms are irrelevant in the scaling limit, as will become clear shortly. To make the dependence on $L$ explicit in the free energy, we perform the substitution 
\begin{equation}
	\bm{x} = L\bm{x'},\quad \bm{\nabla} = L^{-1}\bm{\nabla'},\quad d\bm{x} = L d\bm{x'}.
\end{equation}

The exponent in the integrand of \eqref{eq:partf} can be viewed as the a thermodynamic free energy $\mathcal{F}[h(\bm{x})]$ that is a functional of the random function $h(\bm{x})$. Taking the logarithm of the partition function \eqref{eq:partf}, one gets the statistical mechanical free energy $F$, which in the thermodynamic limit, using saddle-point approximation, can be identified with the thermodynamic free energy evaluated at the height configuration $h^*(\bm{x})$ that maximizes $\mathcal{F}$. After the scaling transformation, it becomes
\begin{equation}
	F\!=\!\int_D\! d\bm{x'}L^2\big(\sigma(\frac{\bm{\nabla'}\phi(\bm{x})}{L})-2(\log q) \phi(\bm{x})\big),
	\label{eq:freeeng}
\end{equation}
where $\phi (\bm{x'})=h^*(\bm{x})$ now satisfy the Lipschitz property of $|\bm{\nabla'}\phi|\le L$ instead, and the integration range becomes the unit disk $D$. From this one can see that in the thermodynamic limit, the higher order terms in the surface tension become irrelevant. The leading Gaussian terms also become irrelevant compared to the linear potential whenever $q\ne 1$. This is because the scaling dimension of $h$ is upper-bounded by 1 due to the Lipschitz condition. In those two cases, it is easy to see that $F$ is minimized at the maximum (for $q>1$) and minimum (for $q<1$) of $\phi(\bm{x})$ respectively, corresponding to a linear and constant scaling of $\phi(\bm{x})$ with system size $L$. While it is clear that the former correspond to $\langle\mathcal{A}\rangle_{\vec{h}}\propto L^2$ and an extensive scaling of entanglement entropy, one might be tempted to conclude that the latter with $\langle\mathcal{A}\rangle_{\vec{h}}=0$, would give rise to sub-area law entanglement. However, this is when the lattice effect becomes important, as even in the minimal height discrete tiling configuration in Fig~\ref{fig:Mintiling}, there is a number (proportional to $L$) of entangled pairs along the bipartition cut, also in agreement with the subleading term in \eqref{eq:Entropy}.

\begin{figure}[t!bh]
	\centering
	\includegraphics[width=0.8\linewidth]{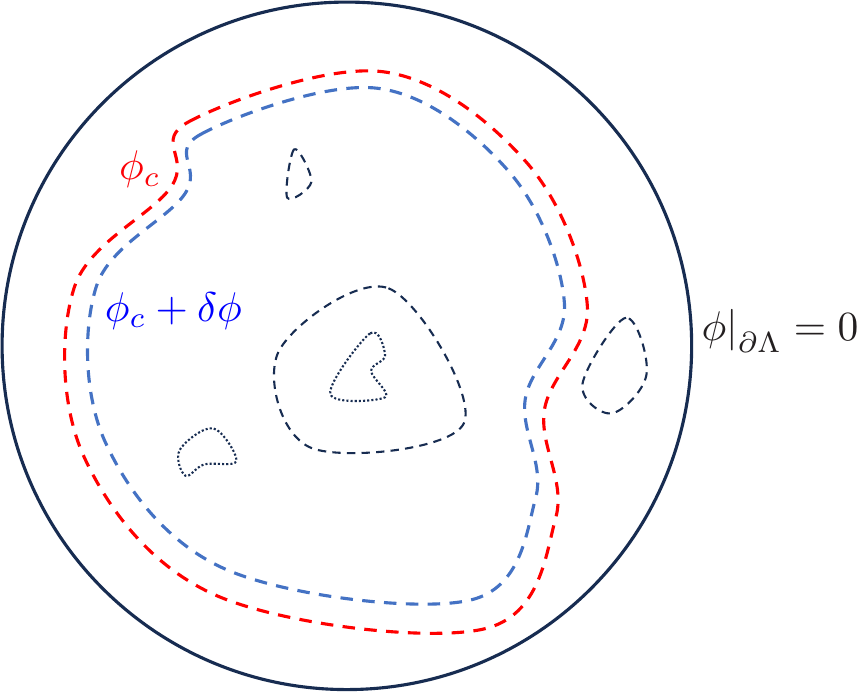} 
	\caption{Topographic view of the height landscape of a typical random surface. The contours of constant height are marked by two different types of lines corresponding to elevation increase and decrease. In a variational argument to decide the most probable incremental change in height $\delta\phi$ in the immediate interior of the red contour of constant height $\phi_c$, we compare the change in surface tension of uniformly elevating the surface inside with the consequential change in entropy of allowing deeper spikes downward inside.}
	\label{fig:topograph}
\end{figure}

At $q=1$, the linear potential in the free energy $F$ disappears, and the competition that determines the scaling of height is now between the surface tension and the hard wall constraint at $h=0$. On one hand, the surface tension favors flatter surfaces with lower height. On the other hand, the hard wall constraint prefers at least slightly higher surfaces so that there is room to fluctuate both above and below, without touching the hard wall. This is precisely the well-studied phenomenon of \textit{entropic repulsion}: The existence of a hard wall pushes the equilibrium surface configuration above its amplitude of fluctuation by a factor of $\sqrt{\log L}$ in dimensions $d\ge 2$~\footnote{Entropic repulsion does not apply to one dimensional systems, where the fluctuation is so large that the surface would not feel the hard wall, see for example Ref.~\cite{1dentrep} and references therein. Nevertheless, intuition of the corresponding scaling there can be gained from the Gumbel universality class of extreme value statistics.}~\cite{Frolich, Bolthausen:1995aa,Bolthausen:2001aa}. An intuitive understanding of the height scaling that balances these two competing effects can be obtained as follows, which we formulate for a general dimensionality $d\ge 2$~\cite{Frolich,Peierls}. Consider a contour as shown in Fig.~\ref{fig:topograph}, along which the height function is a constant $\phi_c$, determined by the competition between the aforementioned two factors for the degrees of freedom outside that contour. Now let the incremental change in height to another contour in the immediate interior be $\delta\phi$. This costs a surface tension contribution of the order $\delta E_c\sim(\delta\phi)^2L^{d-1}$, as the length of the contour scales with one dimension smaller than the dimension of the lattice. So the contribution of all the degrees of freedom enclosed by the contour to the partition function can be expressed as
\begin{equation}
    Z_c(\phi_c)= \int d\delta\phi e^{-\delta E_c-F_i(\phi_c+\delta\phi)},
\end{equation}
where $F_i(\phi_c)$ is the statistical mechanical free energy accounting for the interior of the chosen contour, which depends only on the height along the contour. The main contribution to the partition function of the interior degrees of freedom now come from fluctuations around a flat surface of height $\phi_c+\delta\phi$. In addition to all the fluctuations contained in $e^{-F_i(\phi_c)}$, due to the possibility of having downward spike of depth $\phi_c+\delta\phi$, it comes with an extra factor of
\begin{align}
     e^{-\delta F_i(\phi_c)}=& \frac{e^{-F_i(\phi_c+\delta\phi)}}{e^{-F_i(\phi_c)}}\nonumber\\ \approx& 1+\int_{-\phi_c}^{-\phi_c-\delta\phi}\mathcal{D}\phi_i e^{-\int_D d\bm{x'} \frac{\phi_i^2}{2G_d}}\nonumber\\ \sim & e^{A_c e^{-\phi_c^2/(2G_d)}},
     \label{eq:approx}
\end{align} where $A_c\sim L^d$ is proportional to the area of the region enclosed by the contour, over which area the two-dimensional integral is integrated. $G_d$ is the variance of the Gaussian free field, which is proportional to the correlation function that scales as 
\begin{eqnarray}
G_d\sim
\begin{cases}
\log L, &  d=2\\
O(1), &  d\ge 3.
\end{cases}
\end{eqnarray}
Combining these two factors, the variation in thermodynamic free energy is
\begin{align*}
    \delta\mathcal{F}_c(\phi_c)\approx&\delta E_c+\delta F_i(\phi_c)\\ \sim& L^{d-1}(\mathrm{const.}-Le^{-\phi_c^2/(2G_d)}).
\end{align*}
The saddle point $\delta\mathcal{F}_c=0$ is then achieved at $\phi_c\sim \sqrt{G_d \log L}$. One can verify that the resulting scaling indeed justifies the approximation used in the last line of \eqref{eq:approx}. To conclude the argument, we remark that the same analysis applies irrespective of the shape and location of the contour chosen at the beginning, as long as it is not so far away from the boundary of the system that a downward spike of depth $\phi_c$ inside would violate the Lipschitz property.


The rigorous proof that in the scaling limit, the random surface height scales as $\sqrt{G_d\log L}\propto \log L$ for $d=2$ is more mathematically involved, but can be found in Ref.~\cite{Frolich, Bolthausen:2001aa}, which employs techniques including the large deviation principle, change of measure, and FKG inequalities. From theorem 4 of Ref.~\cite{Bolthausen:2001aa}, follows the $O(L\log L)$ scaling of entanglement entropy at the critical point of our entanglement phase transition (for $s\ge 2$).

Let us now examine the colorless contribution $S_L(q,s=1)$. First, we note that $S_L(q,s=1)<(\log 2) L$. This is an immediate consequence of the fact that there are at most $2^L$ possible height configurations along the cross-section between the two regions. Therefore, we can conclude that for $s\ge 2$ we have:
\begin{eqnarray}
S_L(q,s)=
\begin{cases}
O(L^2), &  q>1\\
O(L \log L), &  q=1\\
O(L), & q<1
\end{cases}
\end{eqnarray}
As stated above $S_L(q,s=1)<(\log 2) L$ for all $q>0$, implying that the area law always holds for $s=1$. 
It is interesting to note that the difference between the colored and colorless  case is most drastic for the $q>1$ phase. In this case the probability is narrowly concentrated near the maximal volume configuration, and surface fluctuations are suppressed. We therefore expect that the contribution to  $S_L(q,s=1)$ from such fluctuations does not scale with system size. 
This is the hallmark of a strong confinement of the fluctuating degrees of freedom to a small region near the center, and would yield a $O(1)$ contribution to the entropy besides the trivial $O(L)$ entropy from cutting horizontal lozenges into halves.
\color{black}
%
\section{Entanglement phase transition in higher dimensions}
\label{sec:higherd}
%
Without constructing explicit models, we make a prediction of how our recipe with the three ingredients of next-neighbor interaction, color matching and boundary condition could result in area violations of entanglement entropy in higher dimensions. Three-dimensional random tilings such as rhombohedra tiling are surveyed in the Ref.~\cite{Henley:1991aa}, along with the behavior of the fluctuation of their height functions in the ``perpendicular space". In three and higher dimensions, the fluctuation of a Gaussian free field are of order 1, which can be seen from a momentum space integration to obtain the Green's function. Conditioned on staying positive due to the boundary condition and next-neighbor interaction, the average height scales as $\sqrt{\log L}$ as a result of the entropic repulsion~\cite{Bolthausen:1995aa}. Therefore, at $q=1$, entanglement entropy of the ground state of such models scales with the $d-1$ dimensional cross-sectional area between subsystems as $L^{d-1}\sqrt{\log L}$. When $q\ne 1$, the linear term in the free energy \eqref{eq:freeeng} still outweighs the gradient term, making the entanglement scaling on the two side of the critical point still obeying area and volume law respectively.

%
\section{Conclusions}
\label{sec:Concl}
%
We presented a generalization of the 1D area law violating model of colored Fredkin spins chain to 2D. The model is built on a platform of fully packed dimers with an internal degree of freedom on a honeycomb lattice. The unique frustration-free ground state exhibits an entanglement phase transition tuned by a deformation parameter $q$ from area law to volume scaling. The entanglement entropy scaling at the critical point $q=1$ can be analyzed with a  Gaussian free field conditioned to be in the positive half-space, rendering an $L \log L$ asymptotic behavior. Analogous though quantitatively different scaling of models in higher dimensions are also discussed.

It is clear that our toy Hamiltonian, despite being completely local, would not be easy to implement experimentally. The most difficult step is to find a mechanism for the next-neighbor interaction and internal color degree of freedom to emerge from the microscopic level. Nonetheless, significant progress has been made in realizing three-body interactions both in cold atom experiments~\cite{Zhou:2022aa}, and in proposals for real materials~\cite{PhysRevB.105.235120}. An even harder obstacle to experimental realization is the gap scaling: in the regime of volume entanglement we expect the gap to vanish very rapidly in the thermodynamic limit, as is the case in the 1D area deformed Motzkin and Fredkin chains~\cite{levine2017gap,Zhang_2017,andrei2022spin}. The scaling of the spectral gap in 2D area law violating systems have been discussed in similar models due to the same authors~\cite{sixnineteen}, which also provide slightly simplified generalizations of the current construction, using 6- and 19-vertex models instead of dimer covering. In the latter realization, the 1D Motzkin chains are also generalized to 2D. How that can be implemented in the lozenge tiling setting however is still unclear. Further simplifications might also be possible in light of the simplification from the three-site Fredkin interaction to the two-site pair-flip interaction studied in Ref.~\cite{Caha2018ThePM}. Last but not least, the volume scaling phase and the critical point might also lead to novel tensor network structures of the ground state, as the one-dimensional Motzkin chains have revealed~\cite{PhysRevB.100.214430,Alexander2021exactholographic}.

\acknowledgments
ZZ thanks Filippo Colomo, Christophe Garban, Vadim Gorin, Yuan Miao, Henrik R{\o}ising and Benjamin Walter for fruitful discussions. ZZ and IK thank Leonid Petrov for discussions. ZZ acknowledges the kind hospitality of the Galileo
Galilei Institute for Theoretical Physics during the workshops ``Randomness, Integrability and Universality'' and ``Machine Learning at GGI". The work of IK was supported in part by the NSF grant DMR-1918207. 

\bibliographystyle{quantum}

\bibliography{Lozenge}

\begin{thebibliography}{10}

\bibitem{Page93}
Don~N. Page.
\newblock ``{Average entropy of a subsystem}''.
\newblock \href{https://dx.doi.org/10.1103/PhysRevLett.71.1291}{Phys. Rev.
  Lett. {\bf 71}, 1291--1294}~(1993).

\bibitem{Hastings_2007}
M~B Hastings.
\newblock ``An area law for one-dimensional quantum systems''.
\newblock \href{https://dx.doi.org/10.1088/1742-5468/2007/08/P08024}{Journal of
  Statistical Mechanics: Theory and Experiment {\bf 2007}, P08024}~(2007).

\bibitem{AnshuEA21}
Anurag Anshu, Itai Arad, and David Gosset.
\newblock ``An area law for 2d frustration-free spin systems''.
\newblock In Proceedings of the 54th Annual ACM SIGACT Symposium on Theory of
  Computing.
\newblock \href{https://dx.doi.org/10.1145/3519935.3519962}{Page 12–18}.
\newblock STOC 2022New York, NY, USA~(2022). Association for Computing
  Machinery.

\bibitem{Calabrese_2009}
Pasquale Calabrese and John Cardy.
\newblock ``Entanglement entropy and conformal field theory''.
\newblock \href{https://dx.doi.org/10.1088/1751-8113/42/50/504005}{Journal of
  Physics A: Mathematical and Theoretical {\bf 42}, 504005}~(2009).

\bibitem{GioevEA06}
Dimitri Gioev and Israel Klich.
\newblock ``{Entanglement Entropy of Fermions in Any Dimension and the Widom
  Conjecture}''.
\newblock \href{https://dx.doi.org/10.1103/PhysRevLett.96.100503}{Phys. Rev.
  Lett. {\bf 96}, 100503}~(2006).

\bibitem{BravyiEA12}
Sergey Bravyi, Libor Caha, Ramis Movassagh, Daniel Nagaj, and Peter~W. Shor.
\newblock ``Criticality without frustration for quantum spin-1 chains''.
\newblock \href{https://dx.doi.org/10.1103/PhysRevLett.109.207202}{Phys. Rev.
  Lett. {\bf 109}, 207202}~(2012).

\bibitem{Movassagh13278}
Ramis Movassagh and Peter~W. Shor.
\newblock ``Supercritical entanglement in local systems: Counterexample to the
  area law for quantum matter''.
\newblock \href{https://dx.doi.org/10.1073/pnas.1605716113}{Proceedings of the
  National Academy of Sciences {\bf 113}, 13278--13282}~(2016).

\bibitem{Zhang5142}
Zhao Zhang, Amr Ahmadain, and Israel Klich.
\newblock ``Novel quantum phase transition from bounded to extensive
  entanglement''.
\newblock \href{https://dx.doi.org/10.1073/pnas.1702029114}{Proceedings of the
  National Academy of Sciences {\bf 114}, 5142--5146}~(2017).

\bibitem{PhysRevB.94.155140}
L.~Dell'Anna, O.~Salberger, L.~Barbiero, A.~Trombettoni, and V.~E. Korepin.
\newblock ``Violation of cluster decomposition and absence of light cones in
  local integer and half-integer spin chains''.
\newblock \href{https://dx.doi.org/10.1103/PhysRevB.94.155140}{Phys. Rev. B
  {\bf 94}, 155140}~(2016).

\bibitem{Salberger:2017aa}
Olof Salberger and Vladimir Korepin.
\newblock ``Entangled spin chain''.
\newblock \href{https://dx.doi.org/10.1142/S0129055X17500313}{Reviews in
  Mathematical Physics {\bf 29}, 1750031}~(2017).

\bibitem{Salberger_2017}
Olof Salberger, Takuma Udagawa, Zhao Zhang, Hosho Katsura, Israel Klich, and
  Vladimir Korepin.
\newblock ``Deformed fredkin spin chain with extensive entanglement''.
\newblock \href{https://dx.doi.org/10.1088/1742-5468/aa6b1f}{Journal of
  Statistical Mechanics: Theory and Experiment {\bf 2017}, 063103}~(2017).

\bibitem{Zhang_2017}
Zhao Zhang and Israel Klich.
\newblock ``Entropy, gap and a multi-parameter deformation of the fredkin spin
  chain''.
\newblock \href{https://dx.doi.org/10.1088/1751-8121/aa866e}{Journal of Physics
  A: Mathematical and Theoretical {\bf 50}, 425201}~(2017).

\bibitem{ArdonneEA04}
Eddy Ardonne, Paul Fendley, and Eduardo Fradkin.
\newblock ``{Topological order and conformal quantum critical points}''.
\newblock
  \href{https://dx.doi.org/https://doi.org/10.1016/j.aop.2004.01.004}{Annals of
  Physics {\bf 310}, 493--551}~(2004).

\bibitem{Kenyon:2008wf}
Richard Kenyon.
\newblock ``Height fluctuations in the honeycomb dimer model''.
\newblock \href{https://dx.doi.org/10.1007/s00220-008-0511-8}{Communications in
  Mathematical Physics {\bf 281}, 675}~(2008).

\bibitem{Roising22}
Zhao Zhang and Henrik~Schou Røising.
\newblock ``The frustration-free fully packed loop model''.
\newblock \href{https://dx.doi.org/10.1088/1751-8121/acc76f}{Journal of Physics
  A: Mathematical and Theoretical {\bf 56}, 194001}~(2023).

\bibitem{Gorin:2021vd}
Vadim Gorin.
\newblock ``Lectures on random lozenge tilings''.
\newblock \href{https://dx.doi.org/DOI: 10.1017/9781108921183}{Cambridge
  Studies in Advanced Mathematics}. Cambridge University Press.
  Cambridge~(2021).

\bibitem{Borodin:2018aa}
Alexei Borodin and Fabio Toninelli.
\newblock ``Two-dimensional anisotropic kpz growth and limit shapes''.
\newblock \href{https://dx.doi.org/10.1088/1742-5468/aad6b4}{Journal of
  Statistical Mechanics: Theory and Experiment {\bf 2018}, 083205}~(2018).

\bibitem{Lamacraft22}
Luke Causer, Juan~P. Garrahan, and Austen Lamacraft.
\newblock ``Slow dynamics and large deviations in classical stochastic fredkin
  chains''.
\newblock \href{https://dx.doi.org/10.1103/PhysRevE.106.014128}{Phys. Rev. E
  {\bf 106}, 014128}~(2022).

\bibitem{qLozenge}
Alexei Borodin, Vadim Gorin, and Eric~M. Rains.
\newblock ``q-distributions on boxed plane partitions''.
\newblock \href{https://dx.doi.org/10.1007/s00029-010-0034-y}{Selecta
  Mathematica {\bf 16}, 731--789}~(2010).

\bibitem{Raghavan:1997vi}
R.~Raghavan, Christopher~L. Henley, and Scott~L. Arouh.
\newblock ``New two-color dimer models with critical ground states''.
\newblock \href{https://dx.doi.org/10.1007/BF02199112}{Journal of Statistical
  Physics {\bf 86}, 517--550}~(1997).

\bibitem{MulticolorDimer}
B.~Normand.
\newblock ``Multicolored quantum dimer models, resonating valence-bond states,
  color visons, and the triangular-lattice ${t}_{2g}$ spin-orbital system''.
\newblock \href{https://dx.doi.org/10.1103/PhysRevB.83.064413}{Phys. Rev. B
  {\bf 83}, 064413}~(2011).

\bibitem{ColorCode}
H.~Bombin and M.~A. Martin-Delgado.
\newblock ``Topological quantum distillation''.
\newblock \href{https://dx.doi.org/10.1103/PhysRevLett.97.180501}{Phys. Rev.
  Lett. {\bf 97}, 180501}~(2006).

\bibitem{Zhang2024bicolorloopmodels}
Zhao Zhang.
\newblock ``Bicolor loop models and their long range entanglement''.
\newblock \href{https://dx.doi.org/10.22331/q-2024-02-29-1268}{{Quantum} {\bf
  8}, 1268}~(2024).

\bibitem{ColorcodeTeo}
Jeffrey C.~Y. Teo, Abhishek Roy, and Xiao Chen.
\newblock ``Unconventional fusion and braiding of topological defects in a
  lattice model''.
\newblock \href{https://dx.doi.org/10.1103/PhysRevB.90.115118}{Phys. Rev. B
  {\bf 90}, 115118}~(2014).

\bibitem{Kesselring2018boundariestwist}
Markus~S. Kesselring, Fernando Pastawski, Jens Eisert, and Benjamin~J. Brown.
\newblock ``The boundaries and twist defects of the color code and their
  applications to topological quantum computation''.
\newblock \href{https://dx.doi.org/10.22331/q-2018-10-19-101}{{Quantum} {\bf
  2}, 101}~(2018).

\bibitem{Vitagliano_2010}
G~Vitagliano, A~Riera, and J~I Latorre.
\newblock ``Volume-law scaling for the entanglement entropy in spin-1/2
  chains''.
\newblock \href{https://dx.doi.org/10.1088/1367-2630/12/11/113049}{New Journal
  of Physics {\bf 12}, 113049}~(2010).

\bibitem{Ramirez_2014}
Giovanni Ramírez, Javier Rodríguez-Laguna, and Germán Sierra.
\newblock ``From conformal to volume law for the entanglement entropy in
  exponentially deformed critical spin 1/2 chains''.
\newblock \href{https://dx.doi.org/10.1088/1742-5468/2014/10/P10004}{Journal of
  Statistical Mechanics: Theory and Experiment {\bf 2014}, P10004}~(2014).

\bibitem{ZHANG2023169395}
Zhao Zhang.
\newblock ``Entanglement blossom in a simplex matryoshka''.
\newblock
  \href{https://dx.doi.org/https://doi.org/10.1016/j.aop.2023.169395}{Annals of
  Physics {\bf 457}, 169395}~(2023).

\bibitem{PhysRevB.106.134420}
Zhao Zhang and Giuseppe Mussardo.
\newblock ``Hidden bethe states in a partially integrable model''.
\newblock \href{https://dx.doi.org/10.1103/PhysRevB.106.134420}{Phys. Rev. B
  {\bf 106}, 134420}~(2022).

\bibitem{PhysRevLett.125.240603}
Giuseppe Mussardo, Andrea Trombettoni, and Zhao Zhang.
\newblock ``Prime suspects in a quantum ladder''.
\newblock \href{https://dx.doi.org/10.1103/PhysRevLett.125.240603}{Phys. Rev.
  Lett. {\bf 125}, 240603}~(2020).

\bibitem{Thurstontileability}
William~P. Thurston.
\newblock ``Conway's tiling groups''.
\newblock \href{https://dx.doi.org/10.1080/00029890.1990.11995660}{The American
  Mathematical Monthly {\bf 97}, 757--773}~(1990).

\bibitem{Chen:2017aa}
Xiao Chen, Eduardo Fradkin, and William Witczak-Krempa.
\newblock ``Gapless quantum spin chains: multiple dynamics and conformal
  wavefunctions''.
\newblock \href{https://dx.doi.org/10.1088/1751-8121/aa8dbc}{Journal of Physics
  A: Mathematical and Theoretical {\bf 50}, 464002}~(2017).

\bibitem{PhysRevB.96.180402}
Xiao Chen, Eduardo Fradkin, and William Witczak-Krempa.
\newblock ``Quantum spin chains with multiple dynamics''.
\newblock \href{https://dx.doi.org/10.1103/PhysRevB.96.180402}{Phys. Rev. B
  {\bf 96}, 180402}~(2017).

\bibitem{cohn2001variational}
Henry Cohn, Richard Kenyon, and James Propp.
\newblock ``A variational principle for domino tilings''.
\newblock \href{https://dx.doi.org/10.1090/S0894-0347-00-00355-6}{Journal of
  the American Mathematical Society {\bf 14}, 297--346}~(2001).

\bibitem{KenyonOkounkov}
Richard Kenyon and Andrei Okounkov.
\newblock ``Limit shapes and the complex burgers equation''.
\newblock \href{https://dx.doi.org/10.1007/s11511-007-0021-0}{Acta Mathematica
  {\bf 199}, 263--302}~(2007).

\bibitem{Destainville}
N~Destainville.
\newblock ``Entropy and boundary conditions in random rhombus tilings''.
\newblock \href{https://dx.doi.org/10.1088/0305-4470/31/29/005}{Journal of
  Physics A: Mathematical and General {\bf 31}, 6123}~(1998).

\bibitem{1dentrep}
Amir Dembo and Tadahisa Funaki.
\newblock ``Stochastic interface models''.
\newblock \href{https://dx.doi.org/10.1007/11429579{\_}2}{Pages 103--274}.
\newblock Springer Berlin Heidelberg. Berlin, Heidelberg~(2005).

\bibitem{Frolich}
J.~Bricmont, A.~El~Mellouki, and J.~Fr{\"o}hlich.
\newblock ``Random surfaces in statistical mechanics: Roughening, rounding,
  wetting,...''.
\newblock \href{https://dx.doi.org/10.1007/BF01010444}{Journal of Statistical
  Physics {\bf 42}, 743--798}~(1986).

\bibitem{Bolthausen:1995aa}
Erwin Bolthausen, Jean-Dominique Deuschel, and Ofer Zeitouni.
\newblock ``Entropic repulsion of the lattice free field''.
\newblock \href{https://dx.doi.org/cmp/1104273128}{Communications in
  Mathematical Physics {\bf 170}, 417--443}~(1995).

\bibitem{Bolthausen:2001aa}
Erwin Bolthausen, Jean-Dominique Deuschel, and Giambattista Giacomin.
\newblock ``Entropic repulsion and the maximum of the two-dimensional
  harmonic''.
\newblock \href{https://dx.doi.org/10.1214/aop/1015345767}{The Annals of
  Probability {\bf 29}, 1670--1692}~(2001).

\bibitem{Peierls}
M.~P. Nightingale, W.~F. Saam, and M.~Schick.
\newblock ``Wetting and growth behaviors in adsorbed systems with long-range
  forces''.
\newblock \href{https://dx.doi.org/10.1103/PhysRevB.30.3830}{Phys. Rev. B {\bf
  30}, 3830--3840}~(1984).

\bibitem{Henley:1991aa}
Christopher~L. Henley.
\newblock ``Random tiling models''.
\newblock \href{https://dx.doi.org/10.1142/9789814503532_0015}{Volume~11 of
  Series on Directions in Condensed Matter Physics, pages 429--524}.
\newblock WORLD SCIENTIFIC. ~(1991).

\bibitem{Zhou:2022aa}
Zhao-Yu Zhou, Guo-Xian Su, Jad~C. Halimeh, Robert Ott, Hui Sun, Philipp Hauke,
  Bing Yang, Zhen-Sheng Yuan, J{\"u}rgen Berges, and Jian-Wei Pan.
\newblock ``Thermalization dynamics of a gauge theory on a quantum simulator''.
\newblock \href{https://dx.doi.org/10.1126/science.abl6277}{Science {\bf 377},
  311--314}~(2022).

\bibitem{PhysRevB.105.235120}
SangEun Han, Adarsh~S. Patri, and Yong~Baek Kim.
\newblock ``Realization of fractonic quantum phases in the breathing pyrochlore
  lattice''.
\newblock \href{https://dx.doi.org/10.1103/PhysRevB.105.235120}{Phys. Rev. B
  {\bf 105}, 235120}~(2022).

\bibitem{levine2017gap}
Lionel Levine and Ramis Movassagh.
\newblock ``The gap of the area-weighted motzkin spin chain is exponentially
  small''.
\newblock \href{https://dx.doi.org/10.1088/1751-8121/aa6cc4}{Journal of Physics
  A: Mathematical and Theoretical {\bf 50}, 255302}~(2017).

\bibitem{andrei2022spin}
Radu Andrei, Marius Lemm, and Ramis Movassagh.
\newblock ``The spin-one motzkin chain is gapped for any area weight
  $t<1$''~(2022).
\newblock  \href{http://arxiv.org/abs/2204.04517}{arXiv:2204.04517}.

\bibitem{sixnineteen}
Zhao Zhang and Israel Klich.
\newblock ``{Coupled Fredkin and Motzkin chains from quantum six- and
  nineteen-vertex models}''.
\newblock \href{https://dx.doi.org/10.21468/SciPostPhys.15.2.044}{SciPost Phys.
  {\bf 15}, 044}~(2023).

\bibitem{Caha2018ThePM}
Libor Caha and Daniel Nagaj.
\newblock ``The pair-flip model: a very entangled translationally invariant
  spin chain''~(2018).
\newblock  \href{http://arxiv.org/abs/1805.07168}{arXiv:1805.07168}.

\bibitem{PhysRevB.100.214430}
Rafael~N. Alexander, Amr Ahmadain, Zhao Zhang, and Israel Klich.
\newblock ``Exact rainbow tensor networks for the colorful motzkin and fredkin
  spin chains''.
\newblock \href{https://dx.doi.org/10.1103/PhysRevB.100.214430}{Phys. Rev. B
  {\bf 100}, 214430}~(2019).

\bibitem{Alexander2021exactholographic}
Rafael~N. Alexander, Glen Evenbly, and Israel Klich.
\newblock ``Exact holographic tensor networks for the {M}otzkin spin chain''.
\newblock \href{https://dx.doi.org/10.22331/q-2021-09-21-546}{{Quantum} {\bf
  5}, 546}~(2021).

\bibitem{10.21468/SciPostPhysCore.6.3.054}
Henrik~Schou Røising and Zhao Zhang.
\newblock ``{Ergodic Archimedean dimers}''.
\newblock \href{https://dx.doi.org/10.21468/SciPostPhysCore.6.3.054}{SciPost
  Phys. Core {\bf 6}, 054}~(2023).

\end{thebibliography}


\appendix

\section{Ergodicity of the Hamiltonian and uniqueness of the ground state}
\label{sec:ergo}

\begin{figure}[t!bh]
	\centering
	\includegraphics[width=0.8\linewidth]{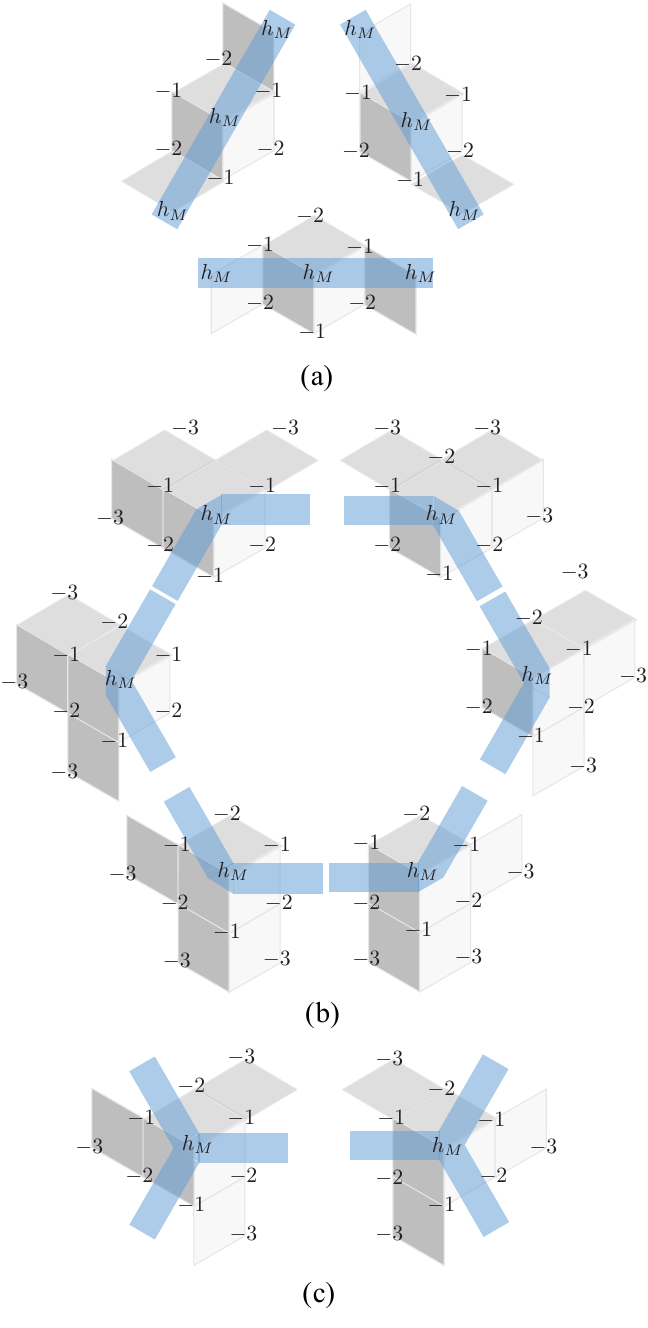} 
	\caption{(a)The three next neighbor configurations that forbid the local maximal height $h_M$ to be lowered, corresponding to two next neighbors along the same axis having the same maximal height. The heights of surrounding sites are marked relative the $h_M$. (b) The six local configurations corresponding to the turning point along the boundary of a plateau of maximal height. (c) Two additional local configurations that allow the maximal height $h_M$ to be lowered even if it's neither isolated nor on the boundary of a plateau. }
	\label{fig:ergo}
\end{figure}

The ergodicity of the Hamiltonian can be established in two steps: First by showing the tiling Hamiltonian $H_T$ relates all color-blind tiling configurations to the rhombille tiling in Fig.~\ref{fig:Mintiling}, corresponding to the minimal height configuration. Then since the color mixing Hamiltonian $H_\Gamma$ mixes the coloring of each neighboring pairs in this configuration, following the reversed path of removal of cubes to the rhombille tiling, the distant entangled pairs in the original random dome surfaces can have any allowed coloring with a total of $s^{N_C}$ choices.

The proof of ergodicity of the tiling moves follows closely the strategy used in Ref.~\cite{Roising22}, which works for dimer covering problems on bipartite lattices. Ergodicity of dimer rotation moves on a general Archimedean lattice was studied systematically with a different approach in Ref.~\cite{10.21468/SciPostPhysCore.6.3.054}. Given any tiling configuration, there must be a site on the triangular lattice $\Lambda$ with maximal height $h_M$, which may not necessarily be unique. The heights of its nearest neighbors alternate between $h_M-1$ and $h_M-2$, but the next-nearest neighbors can either have height $h_M$ or $h_M-3$. If one of the next-nearest neighbours has also height $h_M$, we say our maximal height site is part of a plateau (see Fig.~\ref{fig:ergo} (a,b) for illustration of boundaries of such plateau),  otherwise it is an isolated maximum. The isolated maxima are removable by one of the eight moves in \eqref{eq:FredkinMs}. Maxima sites in the bulk of a plateau (i.e. those with all six next-nearest neighbours are at the same height $h_M$) are not removable. Maxima sites on a plateaux boundary are only removable if they are at the corner of a plateau boundary where there is a turn (Fig.~\ref{fig:ergo} (b)), as along a straight line of boundary both sides in the direction of the boundary are not in the right configuration to allow one of the Fredkin moves, since the next-nearest neighbors are both of the same height (Fig.~\ref{fig:ergo} (a)). So, given any boundary of a plateau, we can always reduce the volume of a surface by starting with removing the unit cubes on the (convex) corners of plateaux boundaries, after which new corners will appear, so that the procedure keeps going. The only scenario such a procedure terminates is when the boundary forms a straight line with the plateau extending to the boundary of the lattice. In that case, both sides of the straight line have the same constant height as the boundary, meaning we have arrived at the lowest height configuration.

Now that we have shown all tiling configurations with the same boundary setting are related the the rhombille tiling, the unique ground state of the Hamiltonian, which first of all, has to comply with that boundary condition, must be a superposition of all the tiling configurations weighted by the volume under the surface, in order to be annihilated by all the projection operators enforcing weighted pairing between locally different configurations.
 
\end{document}